# Plexus: An Interactive Visualization Tool for Analyzing Public Emotions from Twitter Data

**Xiaodong Wu**
Simon Fraser University
Surrey, BC CANADA
xavierw@sfu.ca

**Lyn Bartram**
Simon Fraser University
Surrey, BC CANADA
lyn@sfu.ca

**Chris Shaw**
Simon Fraser University
Surrey, BC CANADA
shaw@sfu.ca

## ABSTRACT
Social media is often used by researchers as an approach to obtaining real-time data on people's activities and thoughts. Twitter, as one of the most popular social networking services nowadays, provides copious information streams on various topics and events. Mining and analyzing Tweets enable us to find public reactions and emotions to activities or objects. This paper presents an interactive visualization tool that identifies and visualizes people's emotions on any two related topics by streaming and processing data from Twitter. The effectiveness of this visualization was evaluated and demonstrated by a feasibility study with 14 participants.

## INTRODUCTION
As of August 2013, there are more than 500 million Tweets posted a day, which means about 5,700 Tweets per second [5]. The contents from social media can be used as an approach to analyzing the patterns of languages. Textual data from Twitter streams contains uncountable amount of valuable information for lots of areas. By developing visualization of the textual data retrieved from thousands of tweets, we can see not only simple attributes like word frequency, but also the underlying connections of different topics and opinions.

The goal of this project is to design and to create a visualization tool using Java to present the semantic analysis results of a large amount of tweets with two topics inputted by a user. The idea is to categorize and to display tweets that are the most related to the emotions in a creative and efficient way so that users can understand the connections between these topics and the main subject. This project will be also beneficial for researchers who are interested in researching and predicting public opinions from social media.

After retrieving data via the official Twitter developer API, the contents will be processed and analyzed and the visualization will display the real-time analysis results as the program is running.

## RELATED WORK
In recent years, social networks have become a hot field for statistician, visual analysts, finance researchers etc. There are lots of research explorations of analyzing and visualizing data from Twitter. The following two fields are closely connected to the scope of our study.

### A. Topic Alignment
Topic detection and modelling are the first step for many Twitter visualizations. Malik et al had created *TopicFlow*, a toolkit for statistical topic modelling and alignment [7]. It can detect the topics and present them as binned topic models. *TopicFlow* was designed to discover the evolution of topics over time from Twitter streams. Hussain et al [2] use power-law graphs to connect all the topics in semantic nets. The node set of a graph is divided into power and non-power nodes and then they are grouped into multiple clusters that are connected to others.

### B. Knowledge Visualization
Our research project fundamentally revolves around information visualization. Liu et al [6] built a knowledge graph by extracting the information structure from Twitter stream and added a semantic dimension. Trees are created and added by connecting the mapped nodes of the knowledge graph.

## MODULES
*Plexus* is essentially written with three main modules to produce a complete visualization.

### 1. Data Processing
*Twitter4J* [15], an unofficial Java library for the Twitter API, processes all the search queries from Twitter.

### 2. Emotion Analysis
Johnson-laird and Oatley [3] produced a theory that there are five basic emotion modes: happiness, sadness, fear, anger, and disgust, and other emotional states are only subsets of these five emotions. Because all emotions are based on the five basic families of emotion modes, this theory states that a word can attribute to more than one emotion modes. Therefore, categorizing a word or text to a single emotion is oftentimes inappropriate. For texts with complex emotions, a quantitative analysis for all five emotion modes is significant for the accuracy of the investigation.

The emotion analysis module in this software is based on the AlchemyLanguage service from IBM [12], which offers a collection of functions and APIs for natural language processing and analysis. The Emotion Analysis function can analyze any textual information in English and return the detected emotions and their relevance values. It can not only analyze the overall emotions of an entire text, but also find out the emotions associated with individual entities or keywords which are also identified by AlchemyLanguage. The relevance values returned along with five emotions indicate the confidence levels for each of them, making it convenient for developers to quantify the emotions.

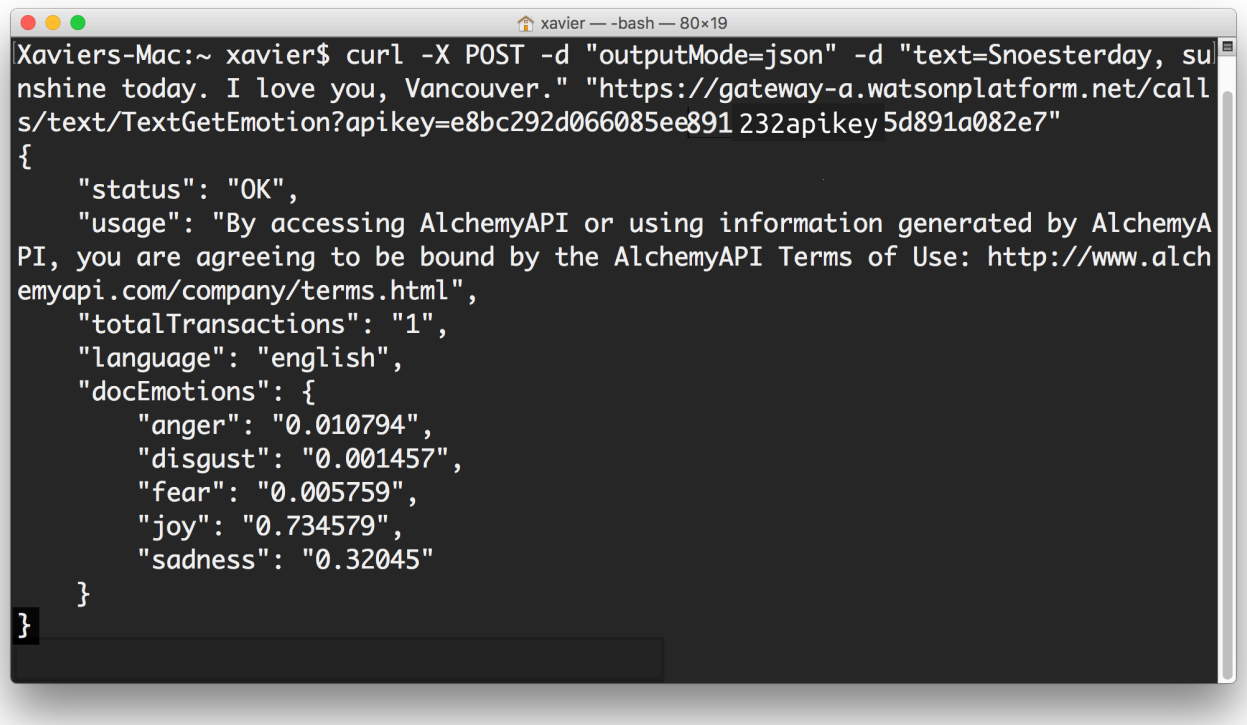

**Figure 1 Emotion Analysis Results for a Simple Tweet from AlchemyLanguage API**

The emotion analysis results returned from AlchemyLanguage are in a *JSON* format, but they are not native *JSON* objects. *JSON-java*, a popular *JSON* package in Java, was used to convert the emotion results to a *JSON* object. This object is parsed to read the values of each emotion and assign them to five *double* variables (*joyValue, angerValue, fearValue, disgustValue, sadnessValue*). A method called *getFinalEmootion* was created in the *GetEmotion* class to return the most relevant emotion of a text based on those five created variables. In this function, a HashMap was created to store the names and values of emotions. I found HashMap is probably the easiest way to find the one with the maximum value out of five variables. It can be used to find either the maximum value or the key with the maximum, or both. If I only need to find the maximum of five values, I can do a multiple comparison using *Math.Max* [13]:

```
double maxVar = Math.max(firstVar, Math.max(secondVar,
Math.max(thirdVar, Math.max( fourthVar, fifthVar)));
```

### 3. Visualization

There are many software toolkits and libraries available for visualization in Java, including *Prefuse*, *Piccolo2D*, *JUNG*, *Graphviz*, and *Gephi*. Eventually *GraphStream* was chosen and used to create interactive visualizations for this project. *GraphStream* is an open-source Java library for modeling and visualizing data in the form of dynamic graphs. It shows dynamic interaction networks for various types of data and handles the graph evolution in time [10]. The changes and transformations of the network can be observed while the program is modeling and visualizing data.

## SOFTWARE DESIGN

Java, the programming language used for this visualization tool, is an objective-oriented language. The software design was expected to accomplish a high cohesion and low coupling among the fundamental modules of *Plexus*. Three classes were created to perform the visualization tasks, along with two other property files.

### 1. TwitterVis

*TwitterVis* is the main class in this software package. The keywords in the Twitter API search query are defined in this class. *TwitterVis* retrieves the data according to the given keywords and pass them to the *GetEmotion* class for emotion analysis.

### 2. GetEmotion

*GetEmotion* submits streamed tweets to the AlchemyAPI service and utilizes a HashMap function to find the emotion mode with the largest relevance value.

### 3. Greetings

*Greetings* provides a simple GUI for users to input two related topics that they are interested in learning through this visualization. The visualization module will initiate after users hit the “submit” button.

### 4. TwitterDevApiKey

*TwitterDevApiKey* stores all the credential information, i.e. API tokens and keys, in the program. The reason for this design is that putting all keys in the same location makes it convenient to organize and update the license information in this package.

### 5. Stylesheet (CSS)

The stylesheet file was created to establish a styling framework for this visualization. All elements’ properties are defined here. Separate styles for users’ actions, such like clicking a node, can also be declared here. The entire stylesheet was written in CSS so that styles of HTML elements can be directly adapted into *Plexus*, a Java program.

## VISUALIZATION DESIGN

A basic goal of visualizing information is to tell a story or solve a problem [14]. To create an effective visualization, several factors such as spatial position need to be considered. *Plexus* provides a simple 2D model to have the visualization results presented as a force-directed network graph, in which two clusters are embedded. From the perspective of visualization, force-directed algorithms create graphs that are aesthetically engaging [9] and interactively intriguing. Since the goal of this visualization design is to create a social media data visualization for users without much prior experience or knowledge in this field.

The biggest challenge for the force-directed approach is that it does not perform well as the number of vertices increases. In other words, producing force-directed algorithms tends to be slow and laggy, and the layouts generated are satisfactory. Another issue that occurred in this project is that the details of individual nodes are difficult to be read as the density of each cluster increases. This can be solved by adjusting the size of nodes in the stylesheet. However, the layout would get messier this way because the algorithm used to produce the force-directed graph calculated the distances between nodes, i.e. the lengths of edges, automatically. Striking a balance of information content and the readability of details is a research problem that occurs to us often when designing the visualization.

### 1. Structure

The initiative of this project was to build a visualization tool that is intuitively simple for users to use and understand without reading values and charts. The visualization is designed to be able to qualitatively present an overview of people's thoughts and reactions to certain topics. Therefore, the structure of data processing is made concise and rational. Figure 2 shows the main framework of *Plexus*, which compares two topics inputted by users, along with several nodes serving as emotions respectively. Each emotion node (the colored circles) has some Tweets whose textual emotion corresponds with the emotion nodes.

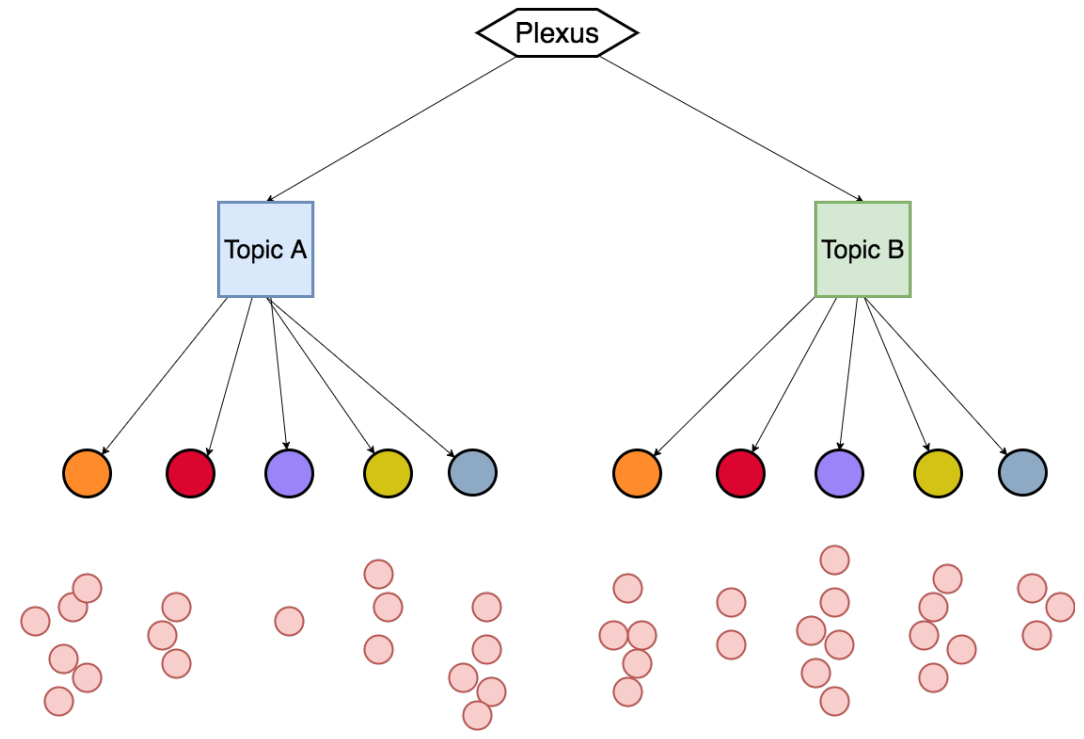

**Figure 3 Structure of Visualization in Plexus**

### 2. Color

Colors for many people are an essential way of presenting information. People's perceptions and interpretations on information are affected by colors. In this visualization, five colors are carefully picked based on several psychologists' [4] and HCI researcher's study [8] to represent the five essential emotions. The connection between each color and emotion was evaluated to be valid by 14 participants. This set of colors is both informational and contrasting so that a user should be able to differentiate between them in the entire visualization.

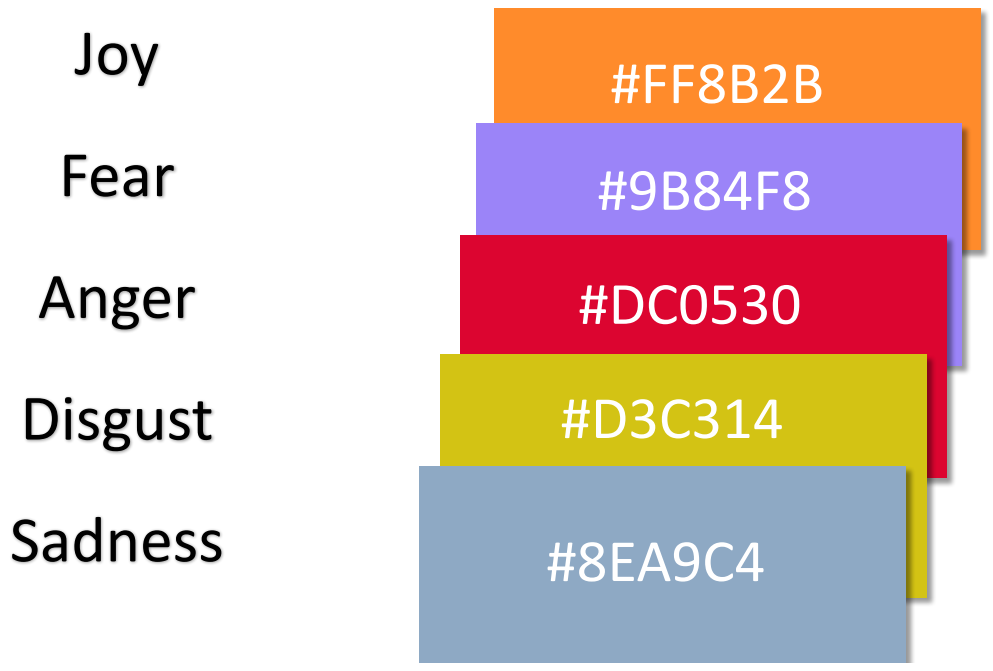

**Figure 2 Colors Picked to Represent the Five Basic Emotions**

The colors picked in Figure 3 were implemented in the visualization and the result is shown in Figure 4.

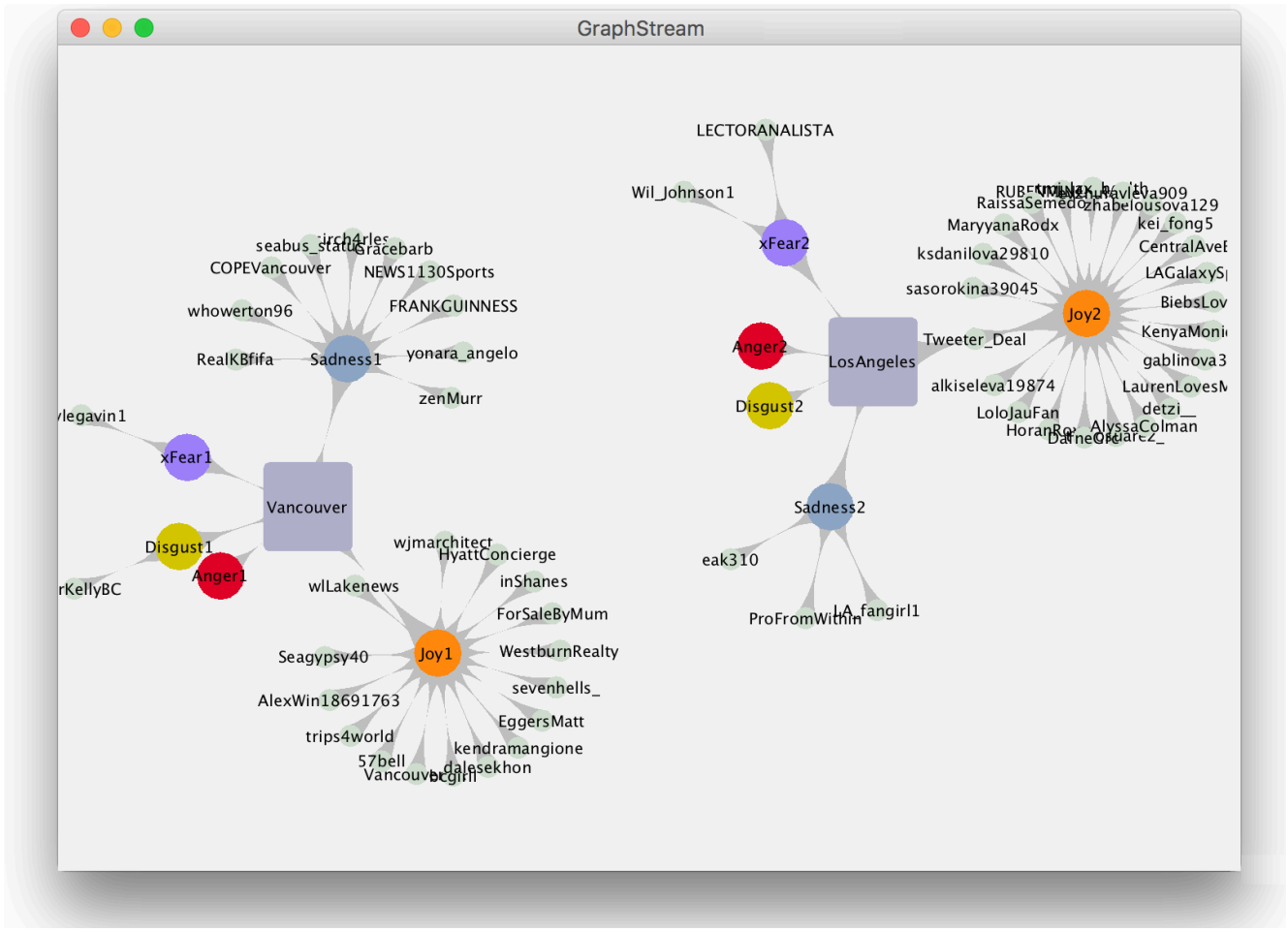

**Figure 4 Visualization Results with Topics "Vancouver" and "Los Angeles"**

### 3. Emoji

Using colors to represent emotions is effective. However, there are still improvement possibilities for emotion representation. Many design techniques, including the fundamental psychological approaches that use dimensional and componential emotion processes [1], are available to server as mediums of the meanings of emotions. An attempt of using Emoji icons was made to improve users' perceptions on emotions. Five icons from Emoji One, an open-source emoji font, were used and modified to adapt to the new design. This new design modification has intuitively made the visualization more explicit. Figure 5 shows how the colors and Emoji icons merge into one element.

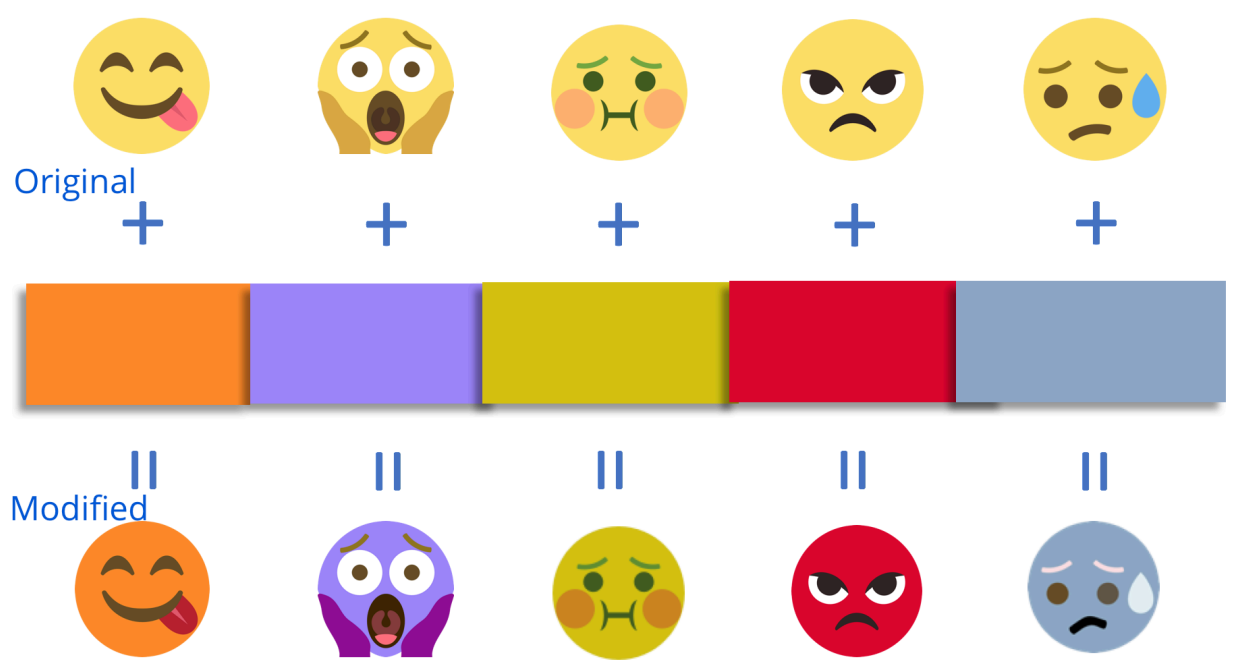

**Figure 5 Redesigned Emoji Icons Based on the Five Picked Colors**

### 4. Elements

The styling details of this visualization are defined in a single CSS stylesheet that is read by the GraphStream library. Many options and properties are available for configuring and customizing the style of visualization elements, which are nodes, edges, and the overall graph.

The improved visualization based on the modified Emoji icons is shown in Figure 6.

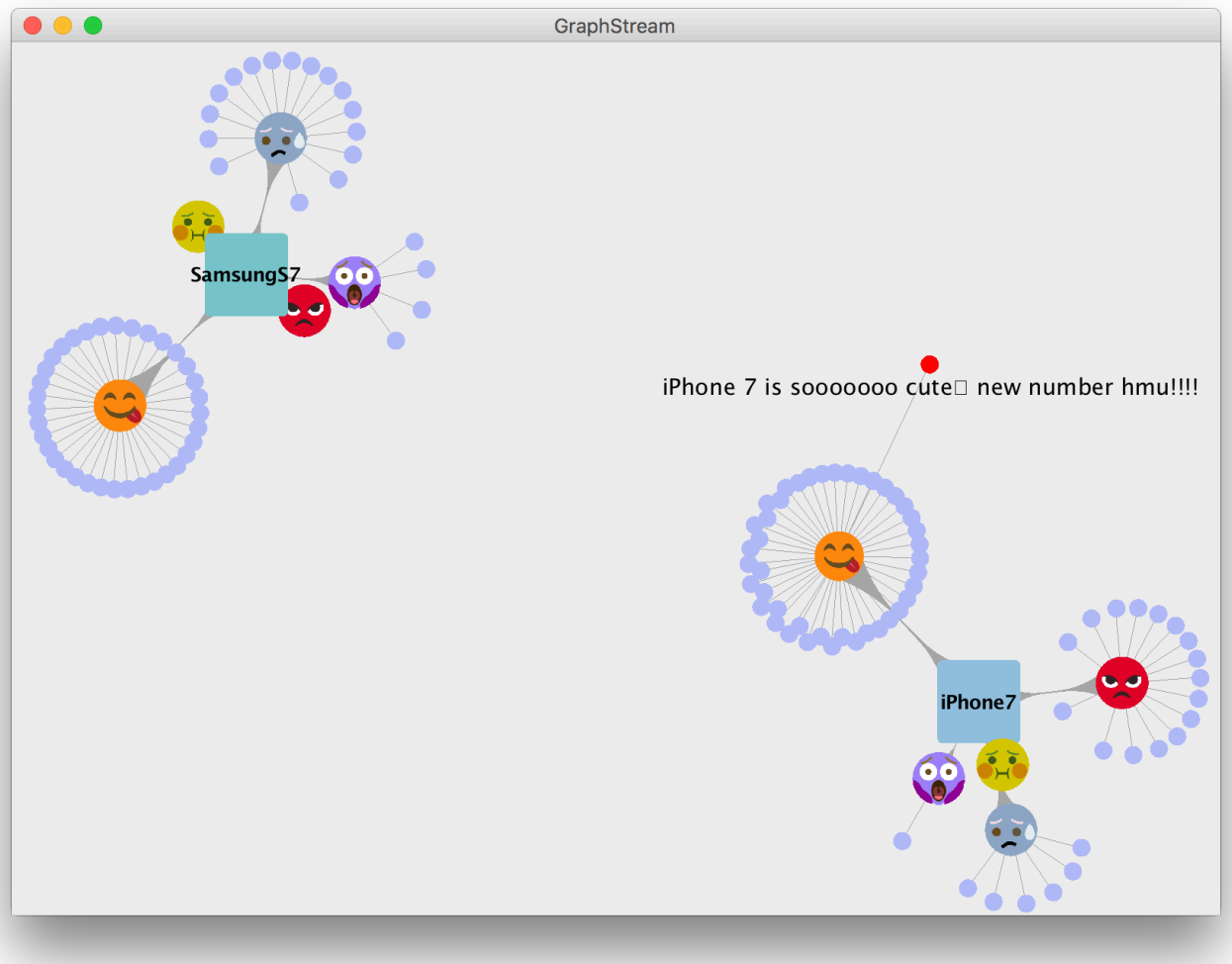

**Figure 6 New Visualization with Emoji Icons Compares People's Emotions to iPhone 7 and Samsung S7**

## FEASIBILITY STUDY

A feasibility study was conducted with 14 participants to understand users' perceptions and comprehensions on the visualization results in *Plexus*. They are all graduate students from the School of Interactive Arts and Technology at Simon Fraser University. Participants were asked to watch a demo video of using *Plexus* and to describe the information they learned after observing the complete visualization process. The usability of this visualization was evaluated through these informal oral surveys. Results show that all of them could understand and describe the meaning of the visualization without much prior explanation. Many of them gave us insightful feedbacks and helpful suggestions. From their suggestions, we will further revise our design of the emoji icons to make the faces more explicit in the context of the dynamic visualization.

## FUTURE WORK

This project has presented an early-stage yet fully functional prototype of an interactive visualization for real-time data from Twitter. There are lots of possibilities for us to explore in terms of data mining, affective computing and visual analytics. The next-stage development of *Plexus* includes two aspects: function and visual design.

### Function

A bountiful amount of information hidden or embedded with tweets is returned every time a data retrieval call happens, including geolocation, language of tweet entries, users' language preferences, and time. *Plexus* can be made more powerful by utilizing these types of information effectively. Several new functions or modules can be added for *Plexus* to make this software more versatile.

*1. Map*

The search operator for geolocalization returns precise results by defining the geocode parameter in the search query [11]. The Twitter Search API will look for Tweets from this defined location, or Tweets created by users from this given location if the first option does not return valid results. Thus, by constructing and embedding a map in *Plexus*, users can investigate the emotions of people from different places to the given topics. For instance, users can explore how the reactions of people from Vancouver and New York City differ in terms of the U.S. presidential election that happened recently.

*2. Time Series*

Processing Twitter data to a time series will enable users to observe how public emotions to specific topics are changing over time. The Tweets retrieved will be categorized by dates and then get processed. A time axis will be added at the bottom of the graph to be manipulated easily by dragging the arrow to choose the year that a user is interested in inspecting.

### Visual Design

This visualization can be made more user-friendly after improving the design for the elements, including the shapes of different types of nodes and the colors of edges.

## CONCLUSION

The overall motivation for this research is to design a visualization tool to solve the problem of information overloading on social media. Our study is primarily an attempted exploration of visual analytics based on social network services. The tool we developed, *Plexus*, managed to foster average users' engagement in learning about topics that they are interested in by using Twitter data. We proposed and realized a functional visualization tool successfully and the results we got from it are significant for us to understand people's reactions to certain topics.